\documentclass[
aps,%
12pt,%
final,%
notitlepage,%
oneside,%
onecolumn,%
nobibnotes,%
nofootinbib,
superscriptaddress,%
noshowpacs,%
centertags]%
{revtex4}
\usepackage[russian]{babel}
\usepackage{epsfig}

\begin{document}
\title{The Asymmetry Coefficient for Interstellar Scintillation
of Extragalactic Radio Sources}

\affiliation{Pushchino Radio Astronomy Observatory, Astro Space
Center, Lebedev Physical Institute, Pushchino, Russia}

\author{\firstname{V.I.}~\surname{Shishov}}

\author{\firstname{T.V.}~\surname{Smirnova}}

\author{\firstname{S.A.}~\surname{Tyul'bashev}}

\received{June 23, 2004}
\revised{September 20, 2004}

\begin{abstract}
Comparing the asymmetry coefficients $\gamma$ and scintillation indices $m$
for observed time variations of the intensity of the radiation of extragalactic
sources and the predictions of theoretical models is a good test of the
nature of the observed variations. Such comparisons can be used to determine
whether flux-density variations are due to scintillation in the interstellar
medium or are intrinsic to the source. In the former case, they can
be used to estimate the fraction of the total flux contributed by the compact
component (core) whose flux-density variations are brought about by
inhomogeneities in the interstellar plasma. Results for the radio sources
PKS~0405--385, B0917${+}$624, PKS~1257--336, and J1819${+}$3845 demonstrate
that the scintillating component in these objects makes up from 50$\%$ to
$100 \%$ of the total flux, and that the intrinsic angular sizes of the sources
at 5~GHz is 10--40~microarcseconds. The characteristics of the medium giving
rise to the scintillations are presented.
\end{abstract}

\maketitle

\section{INTRODUCTION}

Two alternative hypotheses aimed at explaining the phenomenon of
rapid variability of the fluxes of extragalactic sources at centimeter
wavelengths on characteristic time scales of less than a day are discussed
in the literature: (1) intrinsic variability associated with the source and
(2) variability due to interstellar scintillation (see, for example, [1--4]).
Since the velocity of the Earth relative to the interstellar medium displays
seasonal variations with an amplitude of several tens of kilometers, we should
expect variations in the variability time scale over the course of a year in
the case of scintillations. Indeed, such seasonal variations have been observed
for several sources, with the variation time scale increasing appreciably in
the period from August to October [2, 5], in agreement with the decrease in
the velocity of the Earth relative to the interstellar medium in this period.
This demonstrates the interstellar origin of the variability of these sources.

However, for the majority of radio sources, there is no conclusive
proof that fluctuations in their fluxes are due to interstellar
scintillation. In addition, in the case of rapid intrinsic
variability, the angular size of the source must be small enough
that it should inevitably scintillate on inhomogeneities in the
interstellar plasma at centimeter wavelengths. For example, if a
source located at a distance of $10^{28}$~cm displays variability
with a characteristic time scale of the order of $t = 1$~day, its
linear size should be no larger than about $l = 3 \times
10^{15}$~cm, so that its angular size should be no larger than
about $\varphi = 0.1$ microarcsecond (0.1 $\mu$s). This is
appreciably smaller than the angular size of the first Fresnel
zone for the interstellar medium. Consequently, the source will be
point-like from the point of view of interstellar scintillations,
and should accordingly scintillate at centimeter wavelengths, as
pulsars do. Therefore, it is more correct to consider the
following alternatives: (1) intrinsic variability of the source
combined with interstellar scintillation and (2) variability due
only to interstellar scintillation. For sources in which the first
scenario is realized, the problem of distinguishing intrinsic
variations from variations due to interstellar scintillation is
extremely important. We propose here a qualitative test to verify
the nature of observed flux variations by measuring the asymmetry
coefficient of the flux-fluctuation distribution function.

\section{THE ASYMMETRY COEFFICIENT}

When analyzing scintillations of extragalactic radio sources on
inhomogeneities of the interstellar plasma, the most important
measured quantity is the scintillation index:

\begin{figure}[]
\includegraphics[angle=270,scale=0.5]{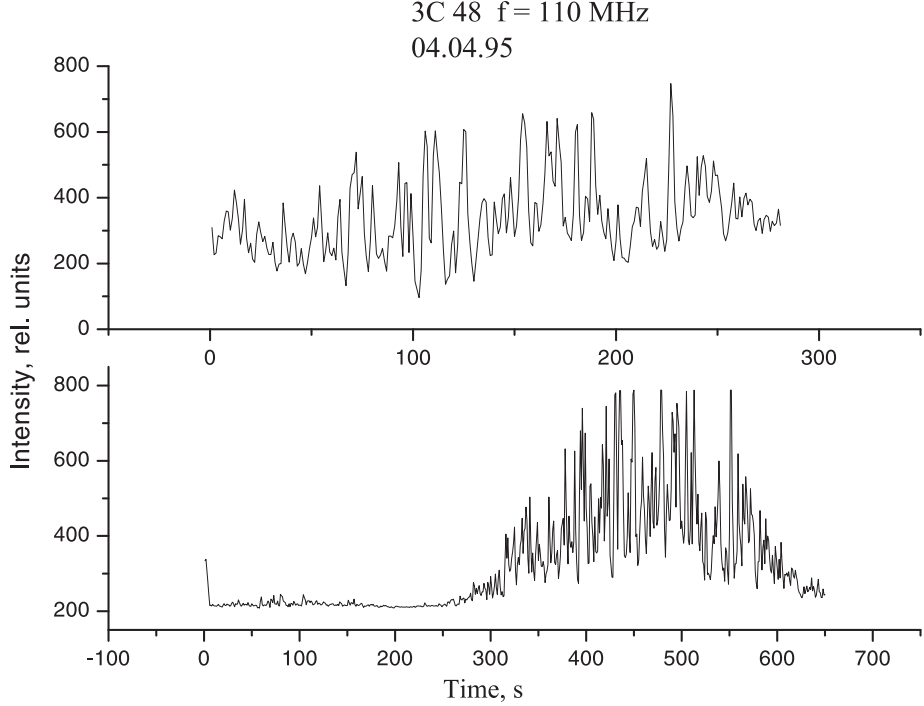}
\setcaptionmargin{5mm}
\onelinecaptionsfalse  
\caption{Original recording of the intensity of the source 3C\,48 as
a function of time at 100~MHz (lower), and the same recording after
eliminating the slow component (upper).
\hfill}
\end{figure}

\begin{gather}
m^2  =  \frac{\langle (I - \langle I \rangle )^2 \rangle}{ \langle
I \rangle^2} , \label{eq:1}
\end{gather}
where $I$ is the measured flux and $\langle I \rangle$ is its mean
value. However, extragalactic sources have complex structures,
consisting of a compact scintillating component and a
non-scintillating extended component. Only the compact component
with angular size of the order of or smaller than the first
Fresnel zone (the ``core'') will scintillate on inhomogeneities of
the interstellar plasma. Consequently, without knowing the flux of
the scintillating component, it is not possible to determine the
corresponding scintillation index, $I_0$:
\begin{gather}
m_0^2  =  \frac{\langle (I - \langle I \rangle )^2 \rangle}{
\langle I_0 \rangle^2}  . \label{eq:2}
\end{gather}

Only the quantity $m_0$ presents physical interest. In~[6--9], it was proposed
to overcome these diffulties associated with observations of scintillating
radio sources using measurements of the asymmetry coefficient of the
flux-fluctuation distribution:
\begin{gather}
\gamma  =  \frac{\langle (I - \langle I \rangle)^3
\rangle}{[ \langle (I - \langle I \rangle )^2 \rangle ]^{3/2} }
               = \frac{M_3}{M_2^{3/2}} .
\label{eq:3}
\end{gather}
Here, $M_3$ and $M_2$ are the third and second central moments of the
flux-fluctuation distribution. Theoretical relations between $\gamma$ and
$m_0$ are known for a number of cases. For example, in the case of
weak scintillations of a point source in the Fraunhofer zone relative to the
outer scale of the turbulence (i.e., the characteristic size of the large-scale
inhomogeneities), the flux-fluctuation distribution follows a Rice--Nakagama
law~[10, 11], and the asymmetry coefficient is given by the relation
\begin{gather}
\gamma  =  \frac{3}{2} m_0 . \label{eq:4}
\end{gather}

\begin{figure}[]
\includegraphics[angle=270,scale=0.5]{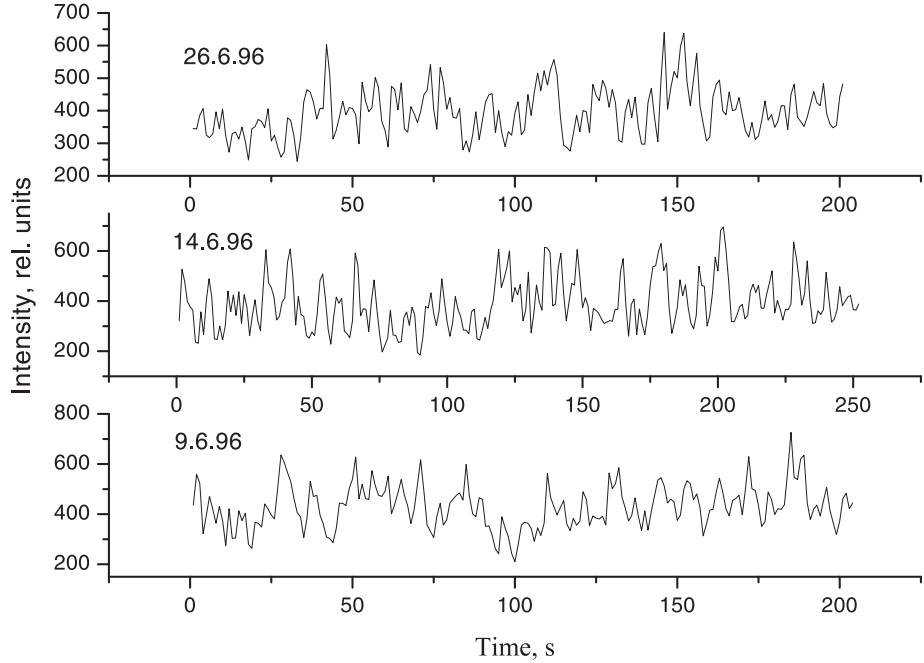}
\setcaptionmargin{5mm}
\onelinecaptionsfalse  
\captionstyle{center} 
\caption{Recording of interplanetary scintillations of 3C\,147 at 110~MHz
over three days of observations. \hfill}
\end{figure}

In the case of weak scintillations of a point source in the
Fresnel zone relative to the inner turbulence scale (i.e., the
characteristic size of small inhomogeneities), this distribution
follows a logarithmic normal law~[12], and
\begin{gather}
\gamma  =  3 m_0
\label{eq:5}
\end{gather}
In the case of a power-law spectrum, a linear dependence between $\gamma$ and
$m_0$ should be preserved:
\begin{gather}
\gamma  =  A m_0 , \label{eq:6}
\end{gather}
where the coefficient $A$ depends on the form of the turbulence spectrum.

This suggestion is supported by the results of Hill et al.~[13, 14], who
carried out numerical computations of flux-fluctuation distribution functions
and determined the second and third moments of the flux fluctuations for the
case of inhomogeneities in the refraction index with a Kolmogorov spectrum
and various inner turbulence scales. These computations yield for a purely
power-law spectrum $A = 2.78$, and for a Kolmogorov spectrum with inner
turbulence scale $l$ equal to the scale of the first Fresnel zone
$r_{\textrm{Fr}}$ ($l/r_{\textrm{Fr}} = 1$) $A = 2.86$. Thus, the coefficient
$A$ increases and approaches the value three as the inner scale for the
turbulence spectrum increases. We can see that, in the case of a Kolmogorov
spectrum, $A$ is close to three, and we can use relation (5) to describe
the relation between $\gamma$ and $m_0$. The computations indicate that (5)
is valid right to values of $m_0$ approaching unity from the weak (unsaturated)
scintillation regime.

We experimentally verified relation (6) using observations of refractive
interstellar scintillations of pulsars at 610~MHz [15], observations of
weak scintillations of the pulsar 1642--03 at 5~GHz [16], and observations
of interplanetary scintillations of the radio sources 3C\,48, 3C\,119,
and 3C\,147 carried out at 110~MH\ using the Large Phased Array of the
Pushchino Radio Astronomy Observatory. In the first set of observations,
the intrinsic variations of the pulsar intensities, with time scales of
several seconds, and variations due to diffractive scintillations, with
time scales of several minutes, were removed by averaging the intensity
over time intervals of about one hour.

Examples of our recordings of the flux variations observed for 3C\,48 and
3C\,147 due to interplanetary scintillations are shown in Figs. 1 and 2.
We used a receiver with a bandwidth of 600~kHz and a time constant of 0.5~s
for these observations. The data were recorded on disk for subsequent
reduction at a rate of 10~Hz. Figure~1 shows an original recording for 3C\,48
together with its noise corridor (lower), as well as the same recording
after filtering out the slow component due to the antenna beam (upper).
Rapid fluctuations of the flux are associated with the passage of the
radiation through the turbulent interplanetary plasma. Figure~2 shows
the flux variations for 3C\,147 observed over three days.

Examples of recordings of flux variations of pulsars due to refractive
scintillations at 610~MHz are presented in~[15]. The results of using these
observations to determine the asymmetry coefficient and scintillation index
are shown in Fig.~3. Refractive interstellar scintillations of a pulsar can
be considered to be weak scintillations of a source whose angular size is
comparable to the scattering angle $\Theta_0$, which is determined by
scattering of the radiation on small-scale (diffractive) inhomogeneities. The
effective size of the inhomogeneities responsible for the refractive
scintillations is the radius of the scattering disk, $L \theta_0$, where $L$
is the effective distance to the turbulent layer. This size is much larger
than the first Fresnel zone $r_{\textrm{Fr}}$, so that, in this case, the
asymmetry coefficient is described by relation (5). The scatter of the points
corresponds to the real statistical errors of the measurements. The turbulence
spectrum for the interplanetary plasma is a power-law over a wide range of
scales, and the power-law index for the three-dimensional spatial spectrum is
close to that for a Kolmogorov spectrum, $n = 11/3$~[17].

\begin{figure}[]
\includegraphics[angle=270,scale=0.5]{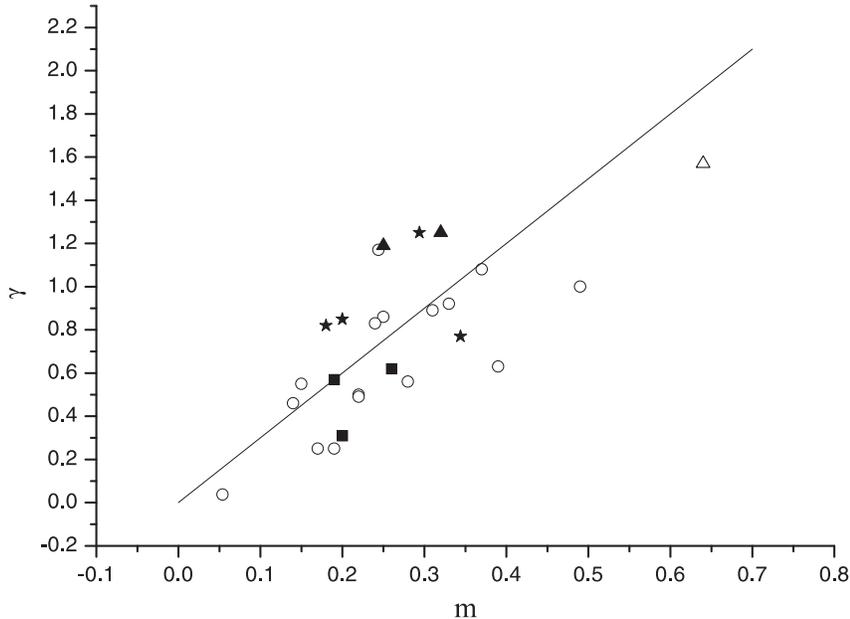}
\setcaptionmargin{5mm}
\onelinecaptionsfalse  
\vspace*{5pt} \caption{Dependence of the asymmetry coeffient $\gamma$ on the
scintillation index $m$. The circles show the data derived from refractive
scintillations of pulsars, the hollow triangles data derived from variations of
PSR~1642--03 at 5~GHz (weak scintillation). The stars, filled triangles and
squares show the data derived from interplanetary scintillations of
3C\,48, 3C\,119, and 3C\,147, respectively. The line corresponds to the
theoretical relation $\gamma = 3 m_0$. \hfill}\vspace*{5pt}
\end{figure}

In contrast to pulsars, which are point sources, the extragalactic objects
we used in our analysis consist of a compact component (core) that scintillates
on inhomogeneities in the interplanetary plasma and a large-scale halo that
does not scintillate. Of course, extragalactic sources can have more complex,
multi-component structures, but for our purposes it is only important that we
divide the source into scintillating and non-scintillating components. To
make the translation from $m$ to $m_0$, we used ratios of the compact and
extended fluxes presented in [18]: $m/m_0 = 0.7$ for 3C\,48 and $m/m_0 = 0.5$
for 3C\,119. We adopted the ratio $m/m_0 = 1$ for 3C\,147. Figure~3 presents
the values of $m_0$ that were used.

Figure~3 shows that the collected data obtained for various media and sources
with various sizes are in good agreement with relation (5). The scatter of
the points about the theoretical curve is only $30\%$, and is primarily
determined by the statistical errors in $\gamma$. Thus, $\gamma$ is a
measurable quantity that can be used to test the hypothesis that interstellar
scintillations are the origin of rapid variations of extragalactic radio
sources.

\section{ASYMMETRY COEFFICIENTS OF A NUMBER OF RAPIDLY VARIABLE
EXTRAGALACTIC RADIO SOURCES}

We used the observed flux variations of the well-known compact radio sources
PKS~0405--385, B0917$+$624, PKS~1257--326, and J1819$+$3845 to determine
the scintillation indices $m$ and asymmetry coefficients $\gamma$ for these
observations. The resulting values of $\gamma$ are listed in the table,
which also gives the scintillation indices $m$ and characteristic scintillation
times $t_0$, with references. The uncertainties in $m$ are $10\% {-} 20\%$.
The characteristic scintillation time was defined as the radius of the
autocorrelation function at the half-maximum level.

\begin{table}
\caption{}
\begin{tabular}{cccccl}
\hline
Источник & Частота наблюдений (ГГц) & m & $t_0$ (час) & $\gamma$ & Ссылка\\
\hline
PKS~0404-385 & 8.64  &  0.08  &  0.41 & +0.12 &  [19]\\
             &  4.8  &  0.11  &  0.55 & +0.62 & \\
             &  2.38 &  0.093 &   1.6 &       & \\
             &  1.38 &  0.063 &   2.6 &       & \\
B0917+624    &  15.0 &  0.01  &   3   & -0.06 &   [3,20]\\
             &   8.3 &  0.02  &   2.4 & +0.22 & \\
             &   5.0 &  0.035 &   7.2 & -0.10 & \\
             &   2.7 &  0.06  &   20  & +0.04 & \\
PKS~1257-326 &  8.6  &  0.05  &  0.27 & +0.18 & [5]\\
             &  4.8  &  0.04  &  0.33 & +0.31 & \\
J1819+3845   &   8.5 &  0.22  &   0.5 & +0.38 & [22]\\
             &   4.8 &  0.29  &  0.53 & +0.78 & [2,21,22]\\
             &   2.2 &  0.24  &       &       & [22]\\
             &   1.3 &  0.13  &   3.5 &       & [22]\\
\hline
\end{tabular}
\end{table}

The dependence of the asymmetry index $\gamma$ on the scintillation index $m$
is shown in Fig.~4, where the data for B0917$+$624 are shown by the hollow
circles, for PKS~1257--326 by the stars, for PKS~0404--385 by the filled
circles, and for J1819$+$3845 by the filled squares. the uncertainties in
$\gamma$ associated with the averaging are roughly $\delta \gamma \cong
0.2$. We can see that the collected points show a well-defined linear
relation between $\gamma$ and $m$, which corresponds to the theoretical
relation (5) with $m =  m_0$; i.e., the case when the flux of the scintillating
component comprises a large fraction of the total flux of the source.

Before turning to a discussion of the data for each source, we will first
present a number of expressions relating the parameters of the scintillations
and those of the medium and source. Data for interstellar scintillations of
pulsars and extragalactic sources shows that the scintillations are weak at
centimeter wavelengths and, in the case of sources with small angular sizes,
the main contribution to the scintillations is made by inhomogeneities with
sizes comparable to the first Fresnel zone, and the spatial scale for the
distribution of the scintillations is comparable to the Fresnel scale:
\begin{gather}
b \cong  r_{\textrm{Fr}}  = \left( \frac{k}{L} \right)^{-1/2} ,
\end{gather}
where $L$ is the effective distance from the observer to the turbulent layer
that is responsible for the scintillations, $k = 2\pi/\lambda$, and $\lambda$
is the wavelength of the observations. If the medium is statistically uniformly
distributed between the source and observer, $L$ corresponds to the distance
between the source and observer, $R$. If the turbulent medium is concentrated
in a fairly narrow layer with thickness $\Delta L \ll R$, then $L$ is the
smaller of the distance from the observer to the layer or the distance from
the layer to the source. Accordingly, the characteristic scintillation time
will be
\begin{gather}
t = \left. \frac{b}{V} = \frac{r_{\textrm{Fr}}}{V} =  \left(
\frac{L}{k} \right)^{1/2} \right/ V,
\end{gather}
where $V$ is the velocity of the Earth relative to the interstellar medium.

The scintillation index for a point source should be
\begin{gather}
m_{0,0} = \left( \frac{f_{cr}}{f} \right)^{\beta},  \qquad \beta =
\frac{n + 2}{4} .
\end{gather}
Here, $n$ is the power-law index of the turbulence spectrum (for a Kolmogorov
spectrum, $n = 11/3$ and $\beta = 1.4$) and $f_{cr} \approx$ 3~GHz is the
critical frequency for the transition from the weak to the strong scintillation
regime [23].

Relations (8) and (9) determine the parameters of scintillations of a source
with small angular size. If the angular size of the source $\varphi_0$ is
larger than the Fresnel angle,
\begin{gather}
\varphi_0 >  2\varphi_{\textrm{Fr}}  = \frac{2}{k r_{\textrm{Fr}}}
= \left( \frac{2}{k L} \right)^{1/2}  ,
\end{gather}
the scintillation index in the weak-scintillation regime is given by the
relation
\begin{gather}
m_0 \cong m_{0,0} \left( \frac{2\varphi_{\textrm{Fr}}}{\varphi_0}
\right)^{\alpha}, \qquad \alpha = \frac{6-n}{2} = \frac{7}{6} ,
\end{gather}
where $m_{0,0}$ is given by (9) and the characteristic scintillation time
is given by
\begin{gather}
t_0 \cong \frac{L \varphi_0}{2 V} .
\end{gather}

When estimating the influence of the angular size of the source on the
scintillation parameters, we should bear in mind that the characteristic
spatial and time scales for the scintillations are determined by the
correlation radius, and correspond to the characteristic radius of
inhomogeneities in the spatial distribution of the intensity fluctuations,
while the angular size of the source is determined by the source brightness
distribution. This is the reason for the additional factor of two in relations
(10)--(12).

\begin{figure}[]
\includegraphics[angle=270,scale=0.5]{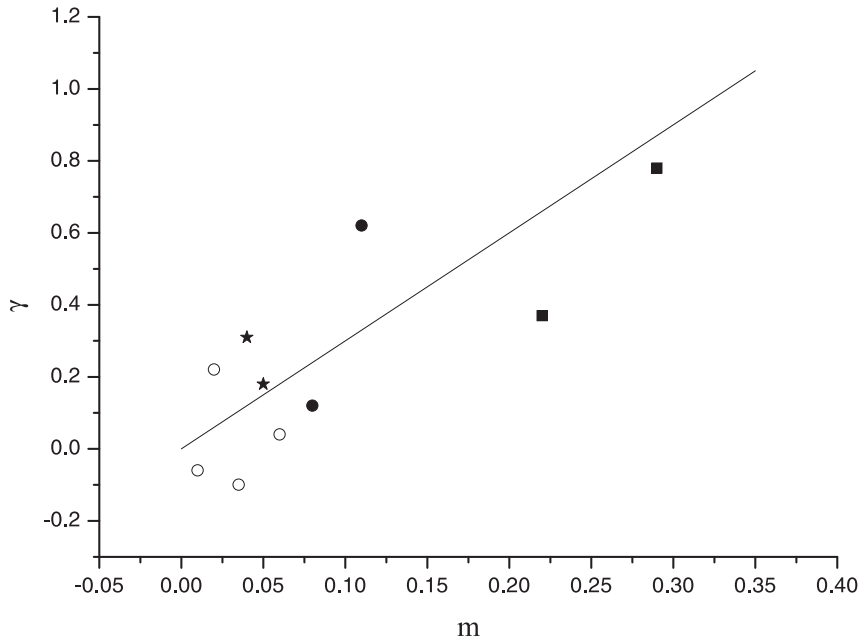}
\setcaptionmargin{5mm}
\onelinecaptionsfalse  
\caption{Dependence of the asymmetry coeffient $\gamma$ on the
scintillation index $m$ for variations of the extragalactic sources.
Shown are the data for PKS~0405--385 (filled circles), B0917$+$624
(hollow circles), J1819$+$3845 (squares), and PKS~1257--326 (stars).
The line corresponds to the theoretical relation $\gamma = 3 m_0$
[Eq. (5)].\hfill}
\end{figure}

Our estimates of $\varphi_{\textrm{Fr}}$ depend appreciably on the effective
distance $L$ to the turbulent layer. Observations show that, at sufficiently
high Galactic latitudes, scintillations of radio sources are determined by
two components of the interstellar medium. The first (component A) is
localized between the spiral arms, and is fairly uniformly distributed in a
thin layer with a thickness of about 1~kpc~[24]. The second (component C)
forms a layer of turbulent plasma with an enhanced level of electron-density
fluctuations located at a distance of about 10~pc from the Sun~[1, 2, 4]. The
relative contributions of these components can vary with direction and
angular size of the source. If the main contribution to the scintillations is
made by the first component, then $\varphi_{Fr,1} \cong 3.5$~$\mu$s at 5~GHz
if $L = 1$~kpc. If the main contribution is made by the second component,
then $\varphi_{Fr,2} \cong 35$~$\mu$s at 5~GHz if $L = 10$~pc. For example,
observations of scintillations of pulsars, whose angular sizes are much
smaller than $\varphi_{Fr,1}$, show that the main effect is due to
inhomogeneities of the first component of the interstellar medium. If the
angular size of the source is much larger than $\varphi_{Fr,1}$, the
main effect will be due to inhomogeneities of the second component of the
interstellar medium. We discuss the data for each source in detail below.

{\textbf{PKS~0405--385.}} The observed flux variations can be understood as
scintillations on inhomogeneities of the local interstellar medium; i.e.,
component C. The values $m \cong 0.11$ and $\gamma \cong 0.62$ at 4.8~GHz
suggest that the scintillating core contains roughly 50\% of the total
flux. The scintillation index for the scintillating component
is $m_0 \cong \gamma/3 \cong 0.2$, so that the scintillations are weak.
The characteristic scintillation time $t_0 \cong 30$~min at 4.8~GHz~[19]
corresponds to the size of the first Fresnel zone [Eq.~(8)] for a distance
$L \cong 10$~pc and a velocity of the observer of $V = 30$~km/s. The
scintillation index decreases at 2.4 and 1.4~GHz, and the characteristic
scintillation time grows rapidly with decreasing radio frequency. This
dependence can be understood as being due to the increase in the apparent
angular size of the source, which is roughly proportional to the square of
the wavelength. In turn, the apparent angular size of the source could be
the result of scattering on inhomogeneities of the interstellar plasma in
a layer that is further from the observer. Studies of scattering of pulsars
have shown that turbulent interstellar plasma with a characteristic thickness
of about 1~kpc is observed in all directions from the Sun [24]. The
characteristic scattering angle in the direction of PKS~0405--385 should
be $\theta_0 \cong 90$~$\mu$s at 2.38~GHz.

{\textbf{B0917}$\pmb{+}$\textbf{624.}} A comparison of the values of $m$ and
$\gamma$ at four frequencies indicates that the scintillating component
comprises close to $100\%$ of the total flux. The small value of the
scintillation index is due not to the flux of the scintillating component
comprising a small fraction of the total, but to the large angular size of
the source. With increasing wavelength $\lambda$, we observe an approximately
linear growth in the scintillation index, $m \propto \lambda$, and in the
scintillation time, $t_0 \propto \lambda$.  These dependences can be understood
only if the angular size of the source appreciably influences the scintillation
parameters, with the apparent angular size of the source being roughly
proportional to the wavelength: $\varphi_0 \propto \lambda$. This dependence
excludes the interstellar medium as the origin of the apparent angular size
of the source, so that this must be the intrinsic angular size. Estimates of
this size depend on the distance to the effective center of the turbulent
layer. We used observations of scintillations of the pulsar B0809$+$74, which
is located close to B0917$+$624 in the sky, to estimate the parameters of the
turbulent medium. This pulsar is located 433~pc from the Sun and has a
velocity $V = 102$~km/s~[25].  The scintillation parameters presented in
[26] correspond to scintillation in a turbulent medium that is uniformly
distributed between the source and observer. Using the scintillation parameters
for B0809$+$74 obtained at 933~MHz, $m = 0.8$ and $t_0 = 2$~hr, together with
Eqs.~(11) and (12), and using the fact that the turbulence spectrum is
Kolmogorov, we obtain $m = 0.08$ and $t_0 = 50$~min at 4.8~GHz. Taking into
account the fact that the pulsar scintillations correspond to the case of a
spherical wave, while the scintillations of the extragalactic source correspond
to the case of a plane wave, and also the fact that the velocity of the
observer is roughly a factor of three lower than the velocity of the
pulsar, we find for a point-like extragalactic source $m = 0.11$ and
$t_0 = 2.5$~hr.  The inferred scintillation parameters of B0917+624 at 5~GHz
correspond to those for an extragalactic radio source with an angular size
of $\varphi \cong 10$~$\mu$s (roughly a factor of three larger than the size
of the first Fresnel zone).

{\textbf{PKS~1257--326.}} The characteristic time scale for the flux
fluctuations shows seasonal variations, providing a direct demonstration
that they have an interstellar origin~[19]. The observed fluctuations can
be understood as scintillations in the local interstellar medium. The values
$m \cong 0.04$ and $\gamma \cong 0.31$  at 4.8~GHz suggest that the
scintillating core contains about $40\%$ of the total flux. The scintillation
index for the scintillating component is $m_0 \cong \gamma/3 \cong 0.10$,
so that the scintillations are weak. The characteristic scintillation time
$t_0 \cong 20$~min~[5], corresponds to the size of the first Fresnel zone for
a distance of $L \cong 5$~pc and a velocity of the observerя $V = 30$~km/s.
The values $m \cong  0.05$ and $\gamma \cong 0.18$ indicate that the
scintillating component comprises 100\% of the total flux at 8.6~GHz, with
the scintillation index being half that for the scintillating component at
4.8~GHz. This corresponds to (11) within the errors in the measured
parameters [according to (11), the decrease in the scintillation index
should be a factor of 2.3].

{\textbf{J1819}$\pmb{+}$\textbf{3845.}} The characteristic time scale for
the fluctuations shows seasonal variations, providing a direct
demonstration that they have an interstellar origin~[2]. The fluctuations
can be understood as scintillations on inhomogeneities of the local
interstellar medium (component C). The values $m \cong 0.29$ and $\gamma
\cong 0.78$ at 4.8~GHz suggest that the scintillating core contains roughly
90\% of the total flux. The scintillation index for the scintillating
component is $m_0 \cong \gamma/3 \cong 0.26$, so that the scintillations are
weak. The characteristic scintillation time, $t_0 \cong30$~min~[2], corresponds
to the size of the first Fresnel zone for a distance of $L \cong 10$~pc and
a velocity of the observer $V = 30$~km/s. The values $m \cong 0.22$ and
$\gamma \cong 0.38$ at 8.5~GHz can be understood as scintillations if the
scintillating component contains about 60\% of the total flux. The
scintillation index begins to fall off at 2.2 and 1.3~GHz, and the
characteristic scintillation time rapidly grows with decreasing frequency.
As for PKS~0405--385, this dependence can be understood as a consequence of
an increase in the angular size roughly in proportion to the square
of the wavelength. The angular size is probably due to scattering on
inhomogeneities in the interstellar plasma in a layer with a thickness of
about 1~kpc. The characteristic scattering angle in this medium in the
direction of J1819$+$3845 should be $\theta_0 \cong 100$~$\mu$s at 2.2~GHz.

\section{DISCUSSION}

Thus, our test of the origin of flux variations of several rapidly variable
extragalactic sources based on comparisons between the asymmetry coefficient
$\gamma$ and the values predicted by scintillation theory demonstrates that
scintillation is the main mechanism giving rise to the flux variations at
frequencies of 8.6~GHz and lower. Comparisons of the measured scintillation
indices and asymmetry coefficients indicate that the scintillating components
comprise from $50\%$ to $100\%$ of the total flux of the sources, so that the
measured scintillation indices are close to those for a one-component
compact source.

The scintillation parameters of the sources correspond to two types of media:
medium~I has a characteristic thickness of about 1~kpc, while medium~II
has a characteristic thickness of about 10~pc. The quasar B0917$+$624
scintillates in medium~I, and indeed, this medium is responsible
for most of the scintillation and scattering of the radio emission of
pulsars~[23]. The angular size of the source grows linearly with growth
in the wavelength.

The scintillations of the remaining three sources occur in medium~II,
which is closer to the observer and has a characteristic thickness of 10~pc.
The parameters of this medium have been estimated in~[1, 2, 4, 5]. At 5~GHz
and higher, the sources have angular sizes comparable to or smaller than the
first Fresnel zone, $\varphi_0 \leq 40$~$\mu$s. The scintillation index at
nearby low frequencies decreases with increasing wavelength, while the
characteristic scintillation time grows roughly in proportion to the square
of the wavelength. This can be understood if radiation that has been scattered
in medium~I is then incident on medium~II; the characteristic scattering
angle should be comparable to the angular size of the first Fresnel
zone (medium II) at 5~GHz. In this case, scintillation in medium~I should
be suppressed by the angular size of the source, which is a factor of five
to ten larger than the angular size of the first Fresnel zone for medium~I.
Overall, a crude estimate of the intrinsic angular sizes of the sources at
5~GHz is $10{\leq} \varphi \leq40$~$\mu$s. More accurate estimates of these
angular sizes require analysis of the temporal structure of the intensity
fluctuations near time lags of 3--10~hr.

We have shown that analyzing the variations of the fluxes of rapidly variable
extragalactic sources incorporating calculations of the asymmetry coefficient
provides an effective method for testing the nature of this variability and
estimating the angular sizes of the sources. The variability in the four
sources we have studied here can be explained as scintillations occuring in two
media: a more extended medium with a thickness of about 1~kpc and the local
interstellar medium near the Sun with a thickness of ${\sim} 10$~pc.  The
scintillations occuring in the extended medium are relatively slow (with
characteristic time scales of roughly several hours), and can be used to
estimate the angular sizes of the sources with resolutions of about
3~$\mu$s. The scintillations occuring in the local interstellar medium are
more rapid (with characteristic time scales of fractions of an hour), and
can be used to estimate the angular sizes of sources with the poorer resolution
of about 30~$\mu$s.

\section{ACKNOWLEDGEMENTS}

This work was supported by the Russian Foundation for Basic Research
(project codes 03-02-16509 and 03-02-16522), the INTAS Foundation (grant
00-849), the National Science Foundation (grant AST 0098685), the Federal
Science and Technology Program in Astronomy, and the program of the Physical
Sciences Branch of the Russian Academy of Sciences ``Extended Objects in the
Universe.''

\end{document}